\documentstyle[aps,prb,preprint,epsf]{revtex}
\begin{document}
\preprint{}
\draft
\title{Spin relaxation of conduction electrons in bulk III-V 
semiconductors}
\author{Pil Hun Song and K. W. Kim}
\address{Department of Electrical and Computer Engineering, North
Carolina State University, Raleigh, North Carolina 27695-7911} 
\date{\today}
\maketitle
\begin{abstract}
Spin relaxation time of conduction electrons through the Elliot-Yafet, 
D'yakonov-Perel and 
Bir-Aronov-Pikus mechanisms is calculated theoretically for bulk 
GaAs, GaSb, InAs and InSb of both $n$- and $p$-type.  Relative 
importance of each spin relaxation mechanism is compared and 
the diagrams showing the dominant mechanism are constructed as a
function of temperature and impurity concentrations.  Our approach 
is based upon theoretical calculation of the momentum relaxation rate
and allows understanding of the interplay between various factors affecting
the spin relaxation over a broad range of temperature and impurity
concentration. 
\end{abstract}



\section{Introduction}

Recently, intensive experimental and theoretical efforts have
been concentrated on the physics of electron spin due to the enormous
potential of spin based devices.  In these so called "spintronic" 
devices,\cite{divi,prin,sham} information is encoded in the spin 
state of individual electrons, 
transferred with the electrons, and finally put under measurement.  Electron
spin states relax, i.e., depolarize, by scattering with imperfections 
or elementary excitations such as other carriers and phonons.  Therefore, 
to realize 
any useful spintronic devices, it is essential to understand
and have control over spin relaxation such that the information is not 
lost before a required operation is completed.    

The investigation of spin relaxation has a long history dating back 
to the fifties and most studies have concentrated on III-V semiconductors 
since direct measurement of spin relaxation time is possible 
through optical orientation in these materials.  Three main spin
relaxation mechanisms, the Elliot-Yafet\cite{elli,yafe} (EY), 
D'yakonov-Perel\cite{dp} (DP) and Bir-Aronov-Pikus\cite{bap} (BAP) 
mechanism have been suggested and confirmed 
experimentally.  Earlier works for spin relaxation are mainly on bulk
systems such as $p$-GaAs,\cite{clar1,fish,maru,zerr} $p$-GaSb,\cite{aron}
GaAlAs\cite{clar2} and $n$-InSb.\cite{chaz}  More 
recently, spin relaxation has also been investigated in quantum well 
structures (GaAs,\cite{ohno} GaAsSb,\cite{hall} InGaAs/InGaAsP\cite{hyla} 
and GaAs/AlGaAs\cite{mali}) as well as in bulk systems 
($n$-GaAs\cite{kikk,zakh} 
and InAs\cite{bogg}).  On the theoretical side, there are recent 
approaches which refine or extend the original calculations of 
Refs.~3 and 4 to explain newly obtained experimental
results.  Flatt\'e and coworkers\cite{bogg,flat} employed a nonperturbative 
14-band calculation for the DP mechanism both for bulk and 
quantum well structures and achieved a better agreement with the 
experimental results.  The BAP process was reconsidered through a 
direct Monte Carlo simulation and extended to quantum wells by 
Maialle and coworkers.\cite{maia}

In the most studies, the strategy has been to find the relevant spin 
relaxation mechanism by comparing experimental results for spin
relaxation time, $\tau_s$, with theoretically predicted dependence 
on temperature or doping concentrations.  Based upon these results, 
a ``phase diagram"-like picture showing the dominant spin 
relaxation mechanism can be constructed
to provide a comprehensive global understanding for competition
of spin relaxation mechanisms.  However, since available experimental 
results for $\tau_s$ are usually limited to a narrow 
range of external physical parameters except some intensively 
investigated materials, such pictures are currently available only 
for $p$-GaAs and $p$-GaSb.\cite{aron}  

In this paper, we calculate the electron spin relaxation time for the EY
($\tau_s^{EY}$), the DP ($\tau_s^{DP}$) and the BAP ($\tau_s^{BAP}$) 
processes for several bulk III-V semiconductors: GaAs, GaSb, InAs and
InSb of both $n$- and $p$-type.  Our result for $\tau_s$ is based 
upon theoretical calculation of the momentum relaxation time,
$\tau_p$.  A diagram is constructed illustrating the dominant spin 
relaxation processes as a function of temperature and impurity 
concentration for each material.  The resulting ``phase
diagrams" for $p$-GaAs and $p$-GaSb are in qualitative agreement with
that of an earlier study.\cite{aron}  The diagrams for the other
materials considered in this work were not available in the literature
and represent an attempt to provide a better understanding of
interplay between various factors for $\tau_s$.  We also discuss some
incomplete aspects of the current theories for spin relaxation.

The rest of this paper is organized as follows.  In Sec.~II the basic
formulation of the three spin relaxation mechanisms is briefly 
described.  The 
details of our calculation for the momentum relaxation time
($\tau_p$) and $\tau_s$ are presented in Sec.~III.   In Sec.~IV the
results for $\tau_s$ are compared with available experimental results
and the ``phase diagrams" for dominant spin relaxation is constructed.
The conclusion follows in Sec.~IV. 

\section{Relevant Spin Relaxation Mechanisms}

\subsection{Elliot-Yafet Mechanism}

The EY mechanism originates from the fact that in the presence of 
spin-orbit coupling, the exact Bloch state is not a spin eigenstate but
a superposition of them.  This induces a finite probability 
for spin flip when the spatial part of electron wavefunction experiences a
transition through scattering even if the
involved interaction is spin independent.\cite{elli,yafe}  The spin 
relaxation time is given by\cite{piku}
\begin{equation}\label{ey}
\frac{1}{\tau_s^{EY}} = A \left(\frac{k_BT}{E_g}\right)^2 \eta^2
\left(\frac{1-\eta/2}{1-\eta/3}\right)^2 
\frac{1}{\tau_p},
\end{equation}
where $E_g$ is the band gap and  $\eta = \Delta/(E_g+\Delta)$ 
with the spin-orbit splitting of the valence band $\Delta$.  $A$ is a 
dimensionless constant and varies from 
2 to 6 depending on dominant scattering mechanism for
momentum relaxation.


\subsection{D'yakonov-Perel Mechanism}

In III-V semiconductors, the degeneracy in the conduction band is 
lifted for ${\bf k} \neq 0$ due to the absence of inversion
symmetry.  The resulting energy difference for electrons
with the same ${\bf k}$ but different spin states plays the role of an 
effective magnetic field and results in spin precession with angular
velocity $\omega({\bf k})$ during the time between collisions.  Since 
the magnitude and the direction of ${\bf k}$ changes in an uncontrolled 
way due to electron scattering with impurities and excitations, this 
process contributes to spin relaxation.  This is called the DP 
mechanism\cite{dp} and $\tau_s^{DP}$ is given by\cite{dp,piku}
\begin{equation}\label{dp}
\frac{1}{\tau_s^{DP}} = Q\alpha^2 \frac{(k_BT)^3}{\hbar^2 E_g} \tau_p,
\end{equation}
where $Q$ is a dimensionless factor and ranges $0.8-2.7$ depending on the
dominant 
momentum relaxation process.  $\alpha$ is the parameter characterizing the 
$k^3$-term for conduction band electrons and is approximately given by
\begin{equation}
\alpha \simeq \frac{4\eta}{\sqrt{3-\eta}}\frac{m_c}{m_{0}}.
\end{equation}
Here $m_c$ and $m_0$ are the effective mass of the conduction electron 
and the electron rest mass, respectively. 

\subsection{Bir-Aronov-Pikus Mechanism}

Electron spin flip transition is also possible by electron-hole scattering
via exchange and annihilation interactions.  This is called the BAP
mechanism and is especially strong in $p$-type semiconductors due to high
hole concentrations.  $\tau_s^{BAP}$ is given by several different
expressions depending on the given external parameters.  In the case of
a nondegenerate semiconductor\cite{bap,piku} ($N_A < N_c$),
\begin{equation}\label{bap1}
\frac{1}{\tau_s^{BAP}} = \frac{2a_B^3}{\tau_0v_B}\left(\frac{2 
\epsilon}{m_c}\right)^{1/2}\left[n_{a,f}|\psi(0)|^4+\frac{5}{3}n_{a,b}
\right],
\end{equation}
where $n_{a,f}$ ($n_{a,b}$) is the concentration of free (bound) holes
and $N_c$ is the critical hole concentration between degeneracy and
nondegeneracy.  $\epsilon$ is the conduction electron energy and $\tau_0$ 
is given by the relation 
$$ \frac{1}{\tau_0} = \frac{3\pi}{64} \frac{\Delta^2_{exc}}{E_B\hbar} $$ 
with $\Delta_{exc}$ the exchange splitting of the exciton ground
state.  $a_B, v_B$ and $E_B$ are defined as
\begin{eqnarray*}
 a_B &=& \frac{\hbar^2\epsilon_0}{e^2m_R} = \left(\frac{m_0}{m_R}\right)
\epsilon_0 a_0; \nonumber \\  
v_B &=& \frac{\hbar}{m_Ra_B}; \nonumber \\
E_B &=& \frac{\hbar^2}{2 m_R a_B^2} = \left(\frac{m_R}{m_0}\right)
\frac{{\cal R}}{\epsilon_0^2},
\end{eqnarray*}
where $m_R$ is the reduced mass of electron and hole, $a_0$ the Bohr
radius ($\simeq$ 0.53 \AA) and ${\cal R}$ the Rydberg constant
($\simeq$ 13.6 eV).  
$\psi({\bf r})$ represents wavefunction describing the relative motion of
electron with respect to hole and $|\psi(0)|^2$ is the Sommerfeld
factor given by
$$|\psi(0)|^2 = \frac{2\pi}{\kappa}(1-e^{-2\pi/\kappa})^{-1}, \ \ \kappa =
\sqrt{\frac{\epsilon}{E_B}}.$$
For degenerate case ($N_A > N_c$), the result is\cite{bap,piku}
\begin{equation}\label{bap2}
\frac{1}{\tau_s^{BAP}} = \frac{2a_B^3}{\tau_0v_B}
\left(\frac{\epsilon}{\epsilon_f} \right) n_a|\psi(0)|^4 \left\{
\begin{array}{ll}
\times (2\epsilon/m_c)^{1/2};\ \ & \mbox{if}\ \ \epsilon_f < \epsilon
(m_v/m_c), \\
\times (2\epsilon_f/m_v)^{1/2};\ \ & \mbox{if}\ \ \epsilon_f > \epsilon
(m_v/m_c), \
\end{array}
\right.
\end{equation}
where $m_v$ is the hole effective mass and $\epsilon_f$ the hole 
Fermi energy, $(\hbar^2/2m_h)(3\pi^2n_a)^{2/3}$.

\section{Calculation}

We first calculate the momentum relaxation time $\tau_p$.  We include 
contributions from the polar optical phonon scattering ($\tau_p^{po}$), 
ionized impurity scattering ($\tau_p^{ii}$), piezoelectric scattering 
($\tau_p^{pe}$), and acoustic phonon deformation potential scattering 
($\tau_p^{dp}$).  Our calculation of $\tau_p$ is performed with three 
simplifying assumptions:\\
(a) the classical Boltzmann statistics is assumed for conduction electrons,\\
(b) the electrons are scattered in a parabolic band,\\
(c) the Mathiessen's rule is applied so that $1/\tau_p = 1/\tau_p^{po}
+ 1/\tau_p^{ii}+1/\tau_p^{pe}+1/\tau_p^{dp}$.\\
Under these assumptions, $\tau_p$ can be obtained in a straightforward
way for the given material parameters of a III-V semiconductor.

According to the Ehrenreich's variational calculation,\cite{ehre1} 
$\tau_p^{po}$ is obtained as
\begin{equation}
\tau_p^{po} = \frac{4}{3\sqrt{\pi}} \frac{\hbar}{\sqrt{{\cal R}k_B T}}
\left(\frac{\epsilon_0
\epsilon_{\infty}}{\epsilon_0-\epsilon_{\infty}}\right)
\left(\frac{m_0}{m_c}\right)^{1/2} \frac{e^{\theta_l/T}-1}{\theta_l/T}
G^{(1)} e^{-\xi},
\end{equation}
where $\epsilon_0$ and $\epsilon_{\infty}$ are the low- and high-frequency 
dielectric constants.  $\theta_l$ is
the longitudinal optical phonon frequency converted in the unit of temperature 
and $G^{(1)} e^{-\xi}$ is calculated as in Ref.~26 as a function of 
temperature and the free carrier density $n$.  

$\tau_p^{ii}$ is described by the Brooks-Herring equation\cite{broo} 
\begin{equation}
\tau_p^{ii} = \frac{1}{3\pi^{3/2}} \frac{\epsilon_0^2/a_0^3}{2N_m + n}
\frac{\hbar (k_B T)^{3/2}}{{\cal R}^{5/2}}
\left(\frac{m_c}{m_0}\right)^{1/2} \int^{\infty}_{0} \frac{x
e^{-x}}{g(n,T,x)} dx,
\end{equation}
where $N_m$ is the concentration of minority impurities, i.e., acceptors for 
$n$-type and donors for $p$-type, and $x$ is a dimensionless quantity
representing $(\epsilon/k_BT)$.  $g(n,T,x)$ is given by
$$ g(n,T,x) = \ln(1+b)-b/(1+b) $$
with
$$ b = \frac{1}{2\pi} \frac{\epsilon_0}{a_0^3 n} \left(\frac{k_B
T}{{\cal R}}\right)^2 \left(\frac{m_c}{m_0}\right) x.  $$

$\tau_p^{pe}$ is given by\cite{meij}  
\begin{equation}
\tau_p^{pe} = \frac{280\sqrt{\pi}}{3} \frac{\hbar}{\sqrt{{\cal R}k_BT}} 
\left(\frac{m_0}{m_c}\right)^{1/2} 
\frac{{\cal R}a_0/e^2}{h_{14}^2(4/c_t+3/c_l)},
\end{equation}
after spherical average of the piezoelectric and elastic constants
over the zinc-blende structure is performed.\cite{zook}
Here $h_{14}$ is the one independent piezoelectric constant and $c_l$
and $c_t$ are the average longitudinal and transverse elastic constants
given by
\begin{eqnarray*}
c_l &=& (3c_{11}+2c_{12}+4c_{44})/5; \nonumber \\
c_t &=& (c_{11}-c_{12}+3c_{44})/5.
\end{eqnarray*}

Finally, Bardeen and Shockley\cite{bard} showed that $\tau_p^{dp}$ is 
given by
\begin{equation}
\tau_p^{dp} = \frac{8\sqrt{\pi}}{3} 
\frac{\hbar{\cal R}^{5/2}}{E_1^2(k_BT)^{3/2}} 
\left(\frac{m_0}{m_c}\right)^{3/2} 
\frac{a_0^3c_l}{{\cal R}},
\end{equation}
where $E_1$ is the deformation potential.

The free carrier concentration $n$ (i.e., electrons for $n$-type and
holes for $p$-type) is calculated from the equation
\begin{equation}
\frac{n(n+N_m)}{N_M-N_m-n} = \frac{N(T)}{2}
\exp(\frac{-E_i}{k_BT})
\end{equation}
Here, $N_M$ is the majority impurity concentration.  $N(T)$ is given by
$[2mk_BT/(\pi\hbar^2)]^{3/2}/4$, where $m$ represents $m_c$ for
$n$-type and $m_v$ for $p$-type, respectively.  $E_i$ is the ionization 
energy for majority impurity and is given by $({\cal R}/\epsilon_0^2) 
(m/m_0)$.  

Table I shows the values of material parameters used in the 
calculation of $\tau_p$ and $\tau_s$.  $E_g(T)$ is obtained by linearly 
interpolating or extrapolating $E_{g,l}$ and $E_{g,h}$ and $N_m$ is 
fixed to $5\times10^{13}$ cm$^{-3}$ in most cases.  Figure 1 plots the
results of mobility calculation, $\mu = (e/m_c)\tau_p$, for $n$-GaAs 
and $n$-InAs.  Good agreement is obtained with the published result 
of Rode and Knight\cite{knig} for $n$-GaAs while our result for
$n$-InAs shows a larger discrepancy up to $\sim$50\% with those of 
Rode\cite{rode}.  This seems to result from the fact that the 
nonparabolicity of conduction band, which we neglected, is stronger 
in InAs.  

Figure 2 illustrates the dominant momentum relaxation mechanism for 
$n$-GaAs as a function of temperature and impurity
concentration.  It is found that the contribution from the polar optical
phonon scattering is dominant for the high-T and lightly-doped regime,
while the ionized impurity scattering dominates otherwise.  The same
qualitative features are found for all other materials investigated,
both for $n$- and $p$-type.

As was noted previously, both $\tau_s^{EY}$ and $\tau_s^{DP}$ include
dimensionless factors, i.e., $A$ in Eq.~(\ref{ey}) and $Q$ in
Eq.~(\ref{dp}), which vary depending on the dominant momentum relaxation 
process.  At the current stage, it is not clear how the crossover 
behavior is given quantitatively when there is a switch between two 
momentum relaxation processes.  
Therefore, we fix the dimensionless constants to their median values,
i.e., $A = 4$ and $Q = 1.75$.  This introduces $\sim$50 \% uncertainty
in our result for $\tau_s^{EY}$ and $\tau_s^{DP}$.  One might correct
this error by directly looking into the dominant momentum relaxation
process.  

To calculate $\tau_s^{BAP}$, we first need to identify the adequate regime
for a given parameter set.  $N_c$ is determined by the
Mott criterion\cite{edwa} $N_c \approx (0.26/a_H)^3$ where $a_H =
a_0\epsilon_0/(m_v/m_0)$.  The thermal averaged value of $1/\tau_s^{BAP}$ 
is obtained as
$$ \langle1/\tau_s^{BAP}\rangle = \frac{2}{\sqrt{\pi}(k_BT)^{3/2}}
\int^{\infty}_{0} \frac{1}{\tau_s^{BAP}(\epsilon)} \sqrt{\epsilon} \
e^{-\epsilon/k_BT} d\epsilon, $$
assuming a classical Boltzmann distribution for conduction electrons.
On the other hand, the expressions for $1/\tau_s^{EY}$ and $1/\tau_s^{DP}$
in Eqs.~(\ref{ey}) and (\ref{dp}) are after thermal averaging with respect 
to $\epsilon$.  A difficulty with the calculation of 
$\tau_s^{BAP}$ lies in the fact that there is no reliable data for 
$\Delta_{exc}$, on which $\tau_s^{BAP}$ has the dependence of 
$\sim 1/\Delta_{exc}^2$, for $p$-InAs and $p$-InSb.  Therefore, we 
examine the tendency of $\tau_s^{BAP}$ as a function of $\Delta_{exc}$ 
as well.

\section{Results and Discussion}

We first compare the relative importance of each spin relaxation mechanism.
Figure~3 shows the dominant spin relaxation processes for $n$-type GaAs, 
GaSb, InAs and InSb.  For $n$-type semiconductors, the contribution
of the BAP mechanism is negligible since the equilibrium hole 
concentration is extremely small.  Therefore, we watch the
competition between the EY and the DP processes.  As shown in Fig.~3, 
it turns out that for all materials investigated there exists a transition 
from the DP-dominant regime to the EY-dominant regime at $T < \sim$5 K 
as the temperature is lowered.  These results are consistent with the 
previously published results that the DP process is the relevant 
spin relaxation mechanism for $n$-GaAs\cite{kikk,flat} and 
$n$-InAs\cite{bogg} at high temperature of 300 K and that the EY 
process is relevant for $n$-InSb at low temperature of T = 1.3 
K.\cite{chaz}  When the acceptor, i.e., the minority
impurity, concentration decreases, we find that the DP-dominant regime 
enlarges.  This can be understood from following consideration.  The 
acceptors in $n$-type materials are always ionized and the
decrease in the acceptor concentration corresponds to the decrease in
the number of scattering centers for ionized impurity scattering
procedure, which is the main momentum relaxation mechanism at low
temperature.  Therefore, a larger $\tau_p$ results as the acceptor
concentration decreases and this induces a larger $\tau_s^{EY}$
and a smaller $\tau_s^{DP}$ since $\tau_s^{EY} \sim \tau_p$ and
$\tau_s^{DP} \sim 1/\tau_p$.  

The diagrams for $p$-type materials are illustrated in Fig.~4 with $10^{14}$ 
cm$^{-3}$ $< N_A <$ $10^{20}$ cm$^{-3}$ and $N_D = 5\times
10^{13}$ cm$^{-3}$.  For $p$-type materials, no systematic 
changes are found when the minority carrier concentration is
varied.  For $p$-GaAs and $p$-GaSb, we find that the BAP (DP) is
dominant in the low-T (high-T) and high (low) doping regime.  This
is in qualitative agreement with the results of Aronov {\it et 
al.},\cite{aron} in which diagrams of the same idea were constructed 
based on experimental results.  For $p$-InAs, a similar feature 
to those of $p$-GaAs and $p$-GaSb is found for 
$\Delta_{exc}$ = 10 eV, and as $\Delta_{exc}$ decreases, the BAP
dominant regime becomes smaller.  For $p$-InSb, we obtain similar results 
to those for $p$-InAs as a function of $\Delta_{exc}$.  Figure~4(d) shows 
the case of $\Delta_{exc} =$ 0.2 eV where a BAP-dominant regime exits at 
$T <$ 100 K and intermediate doping concentrations.  We find abrupt 
discontinuities in $\tau_s^{BAP}$ at $N_A = N_c$, which results in 
unphysical sharp cusps at $N_A \simeq 10^{18}$ cm$^{-3}$ in Fig.~4.  
This is an artifact resulting from the fact that no quantitative 
expression for $1/\tau_s^{BAP}$ is available for the crossover between 
nondegenerate [Eq.~(\ref{bap1})] and degenerate [Eq.~(\ref{bap2})] hole 
regimes.  Experimentally, it was found that there exists an
intermediate regime at $N_A \approx N_c$ where $\tau_s$ remains nearly 
flat with respect to the change in $N_A$ and that the range of such 
intermediate regime varies depending on the material.\cite{aron}

Figures~5 and 6 provide the total spin relaxation time, i.e., $\tau_s = 
(1/\tau_s^{EY} + 1/\tau_s^{DP})^{-1}$ for $n$-type and $\tau_s =
(1/\tau_s^{EY} + 1/\tau_s^{DP} + 1/\tau_s^{BAP})^{-1}$ for
$p$-type, respectively.  $\tau_s$ ranges from 1 ps to 100 ns for 
$n$-type materials and from 0.1 ps to 10 ns for $p$-type materials, 
respectively, over the parameter space shown in Figs.~5 and 6.  
For $n$-type materials, $\tau_s$ increases as $T$ decreases with the 
longest $\tau_s$ found at $N_D \sim 10^{17}-10^{18}$ cm$^{-3}$ 
instead of in purer materials.  This is because the regime shown 
in Fig.~5 is dominated solely by the DP-process and $1/\tau_s^{DP}$, 
which is proportional to $\tau_p$, increases as the impurity 
concentration decreases.  The same qualitative feature has also 
been found in a recent experiment.\cite{kikk}  In our result
for $n$-GaAs, $\tau_s$ ranges from 5 ns to 60 ns for $T$ = 25 K,
which gives a reasonable agreement with the experimental result of
Ref.~19 ($\tau_s \sim$ 70 ns at $T$ = 20 K).  As for $n$-InAs
with $N_D = 10^{16}$ cm$^{-3}$ and $T$ = 300 K, our result gives 
$\tau_s$ = 12 ps which compares very well with a recent experimental 
result of $\tau_s = 19 \pm 4$ ps.\cite{bogg}

At lower temperature, we find a discrepancy with recent experimental
result for $n$-GaAs.  In experiment\cite{kikk} $\tau_s \simeq$ 100 
ns at 5 K for $N_D = 10^{16}$ cm$^{-3}$ was reported, while our 
result predicts a larger value 
of $\tau_s \simeq 6\times10^3$ ns.  Reference.~19 suggested 
that the main spin relaxation at this low temperature regime is due 
to the EY mechanism.  According to our result, however, since $\tau_p
\sim$ 1 ps and $\tau_s^{EY}$ and $\tau_s^{DP}$ are given by $7 \times 
10^4$ ns and $6\times10^3$ ns, respectively, neither the 
EY nor the DP mechanism provides a satisfactory explanation
for the experimental result.  Very recently, in ref.~20, spin
relaxation time of 290 $\pm$ 30 ns at 4.2 K was reported for bound 
electrons to donors in $n$-GaAs and the relevant spin relaxation
mechanism was proposed to be the hyperfine interaction with 
nuclei,\cite{abra} which was not taken into account in our current 
work.  A further research incorporating this effect is 
needed to resolve the discrepancy between our result 
and the experimental result of Ref.~19.

In $p$-type materials, smaller $\tau_s$, i.e., stronger spin relaxation 
rate, than that in $n$-type materials is found due to
the effect of the BAP process.  The strong discontinuities at $N_A = 
N_c$ are also noticeable in Fig.~6 due to the incompleteness
of the BAP expressions given by Eqs.~(\ref{bap1}) and (\ref{bap2}),
as mentioned earlier.

\section{Conclusion}

In this paper, we calculated theoretically $\tau_s$ for several bulk 
III-V semiconductors and compared the contributions from the
three main spin relaxation mechanisms as a function of temperature 
and donor/acceptor concentrations.  In $n$-type materials, the
DP mechanism is found to be dominant down to very low temperature,
below which the EY mechanism dominates.  While our calculated
spin relaxation times are in reasonable agreement with the 
experimental results for high temperature regime of $T 
> \sim$20 K, there exists a discrepancy at $T \sim$ 5 K for 
$n$-GaAs.  Further theoretical efforts incorporating other spin 
relaxation mechanisms neglected in this paper are needed for 
its resolution.  As for $p$-type materials, the BAP (DP) mechanism 
is dominant at low (high) temperature and high (low) acceptor 
concentrations.  We find that the 
crossover between various regimes for spin relaxation requires a
further 
theoretical investigation for a more thorough understanding and 
realistic comparison with experimental data.  This is especially the 
case for the crossover between nondegenerate and degenerate hole 
regimes for the BAP process.\\

\noindent{\bf Acknowledgment}\\

We are thankful to M. I. D'yakonov for critical comment and to 
J. M. Kikkawa for useful discussion.  This work is supported by 
the Office of Naval Research and the Defense Advanced
Research Projects Agency.

\begin{table}
\caption{Material parameters.  $N_c$ is from the relation $N_c
\approx (0.26/a_H)^3$ and all other numbers are from
Ref.~34 unless specified otherwise.}
\begin{center}
\begin{tabular}{l|r|r|r|r}
 & GaAs & GaSb & InAs & InSb \\\hline
$m_c/m_0$ & 0.065 & 0.0412 & 0.023 & 0.0136 \\
$m_v/m_0$ & 0.5 & 0.28 & 0.43 & 0.45 \\
$\Delta$ (eV) & 0.341 & 0.75 & 0.38 & 0.85 \\
$E_{g,l}$ (eV) & 1.52 (0 $K$) & 0.822 (0 $K$) & 0.418 (4.2 $K$) &
0.235 (1.8 $K$) \\
$E_{g,h}$ (eV) & 1.42 (300 $K$) & 0.75 (300 $K$) & 0.354 (295 $K$) &
0.23
(77 $K$) \\
$\epsilon_0$ & 12.515 & 15.69 & 15.15 & 16.8 \\
$\epsilon_{\infty}$ & 10.673 & 14.44 & 12.25 & 15.68 \\
$\theta_l\ (K)$ & 410 & 335 & 343 & 280 \\
$c_{11}$ (dyn/cm$^2)$ & 1.221$\times10^{12}$ & 8.834$\times10^{11}$ &
8.329$\times10^{11}$ & 6.669$\times10^{11}$ \\
$c_{12}$ (dyn/cm$^2)$ & 5.66$\times10^{11}$ & 4.023$\times10^{11}$ &
4.526$\times10^{11}$ & 3.645$\times10^{11}$ \\
$c_{44}$ (dyn/cm$^2)$ & 5.99$\times10^{11}$ & 4.322$\times10^{11}$ &
3.959$\times10^{11}$ & 3.02$\times10^{11}$ \\
$h_{14}$ (V/cm) & 1.45$\times10^7$ & 9.5$\times10^{6}$ &
3.5$\times10^{6}$
& 4.7$\times10^{6}$ \\
$E_1$ (eV) & $6.3^a$ & $6.7^b$ & $4.9^b$ & $7.2^b$ \\
$\Delta_{exc}$ ($\mu$eV) & $50^c$ & $24^d$ & unknown & unknown \\
$N_c$ (cm$^{-3})$ & 7.53$\times10^{18}$ & 6.71$\times10^{17}$ &
2.7$\times10^{18}$ & 2.27$\times10^{18}$
\end{tabular}
\end{center}
$^a$Ref.~35.
$^b$Ref.~33.
$^c$Ref.~36.
$^d$Ref.~12.
\end{table}

\begin{figure*}
\centerline{\epsfxsize=10cm \epsfbox{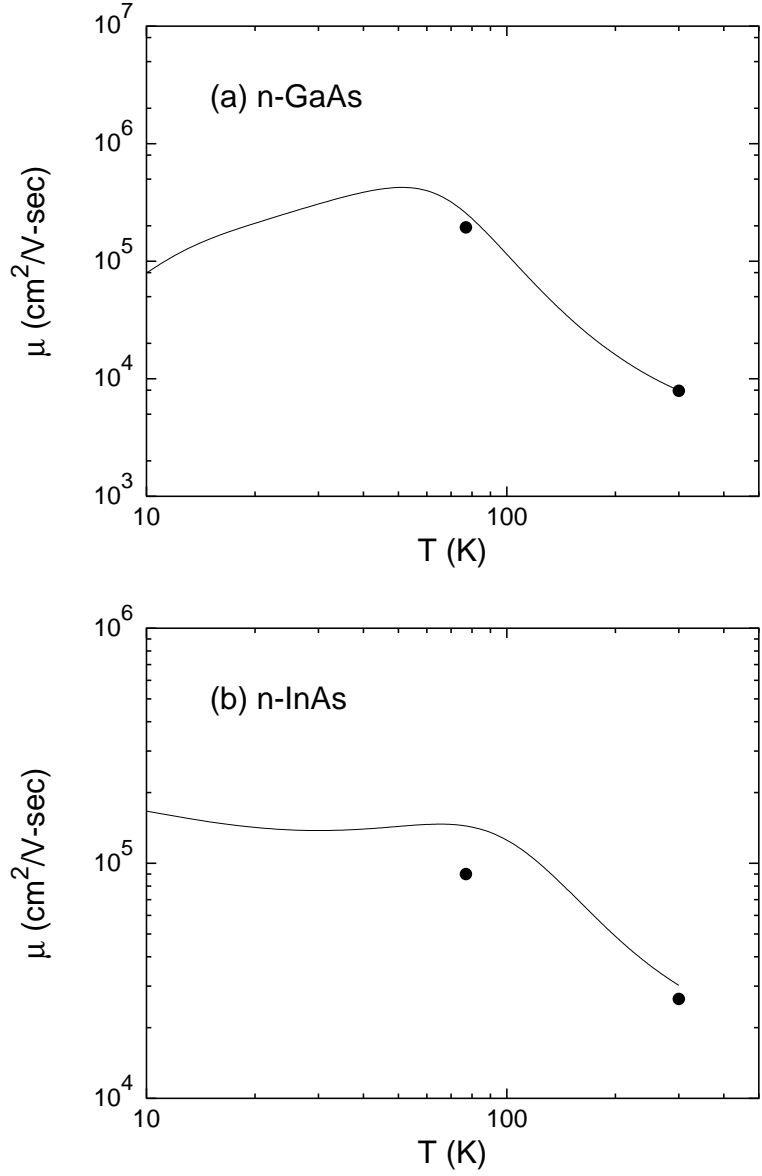}}
\vspace{4mm}
\caption{Mobility vs. temperature for (a) $n$-GaAs for $N_D = 10^{14}$ 
cm$^{-3}$ and $N_A = 5\times10^{13}$ cm$^{-3}$ and (b) $n$-InAs for
$N_D = 2\times10^{16}$ cm$^{-3}$ and $N_A = 5\times10^{13}$ cm$^{-3}$.
The lines are our calculation and the points are from (a) Rode and 
Knight\cite{knig} and (b) Rode.\cite{rode}
}
\end{figure*}
\noindent
\pagebreak
\begin{figure*}
\centerline{\epsfxsize=10cm \epsfbox{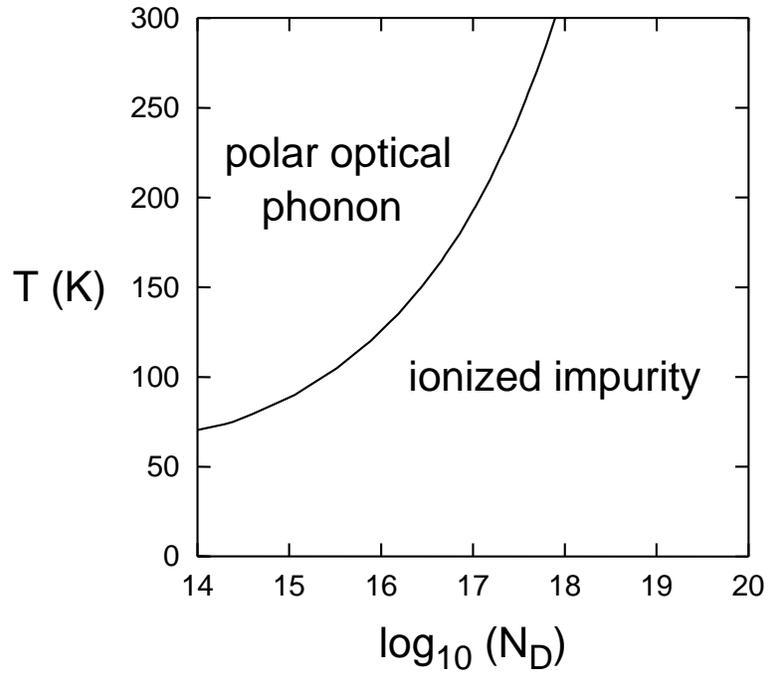}}
\vspace{4mm}
\caption{Dominant momentum relaxation process for $n$-GaAs as a function
of temperature and donor concentration with $N_A = 5\times10^{13}$ cm$^{-3}$.
$N_D$ is in cm$^{-3}$.
}
\end{figure*}
\noindent
\pagebreak
\begin{figure*}
\centerline{\epsfxsize=14cm \epsfbox{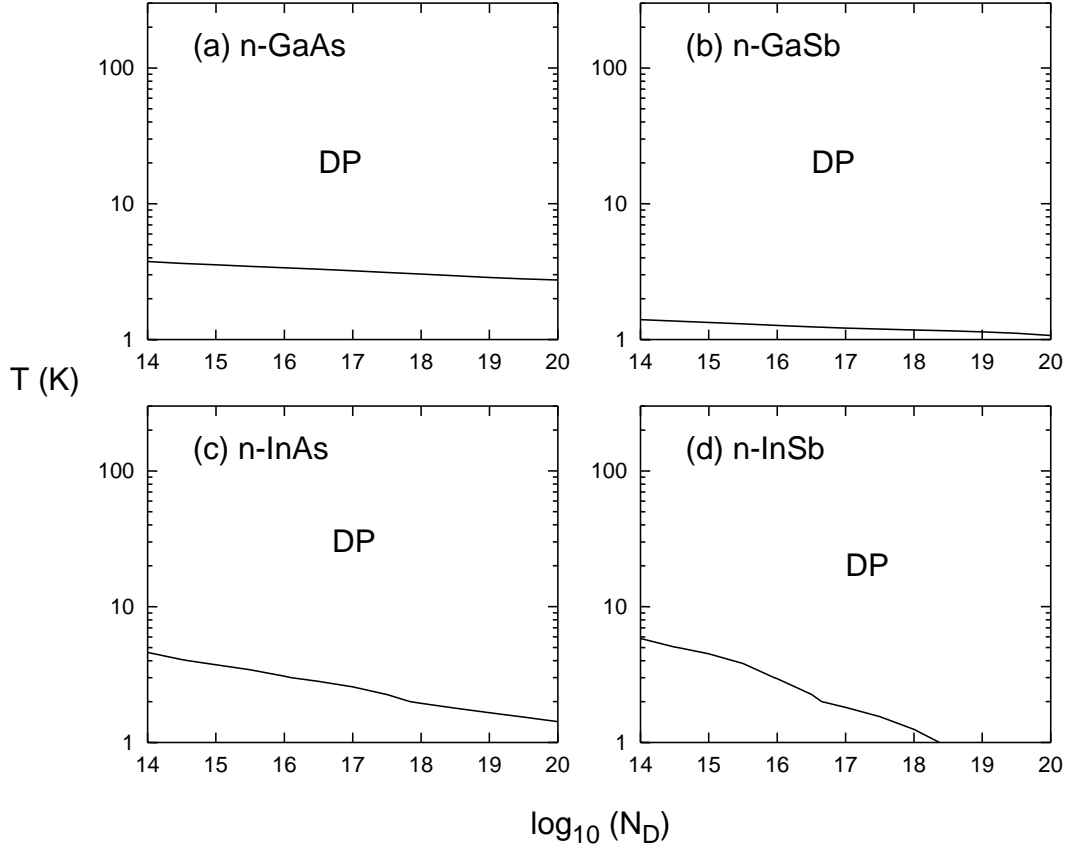}}
\vspace{4mm}
\caption{Dominant spin relaxation mechanism for $n$-type materials.
The higher temperature regime is governed by the DP mechanism as
shown while the lower temperature regime by the EY mechanism.
$N_D$ is in cm$^{-3}$ and $N_A$ is fixed to $5\times10^{13}$ cm$^{-3}$.  
Material parameters are as specified in Table I.
}
\end{figure*}
\noindent
\pagebreak
\begin{figure*}
\centerline{\epsfxsize=14cm \epsfbox{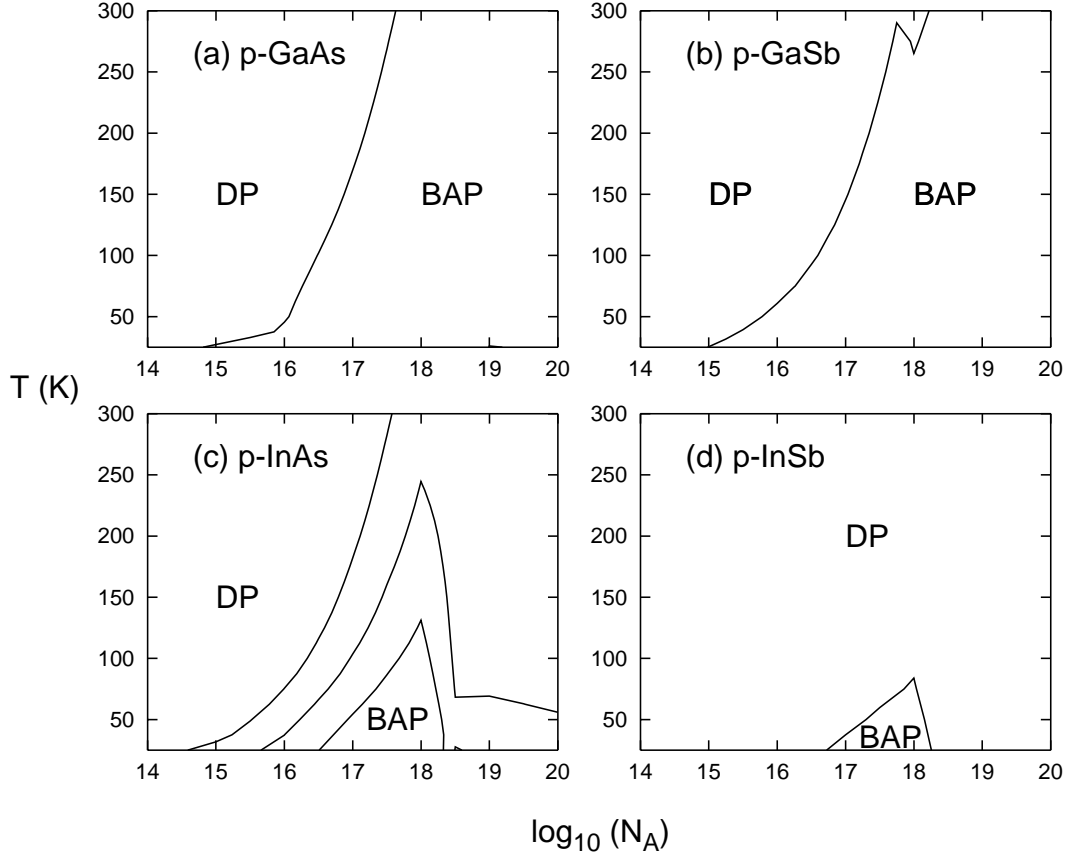}}
\vspace{4mm}
\caption{Dominant spin relaxation mechanism for $p$-type materials.
$N_A$ is in cm$^{-3}$ and $N_D$ is fixed to $5\times10^{13}$ cm$^{-3}$.  
$\Delta_{exc}$ = 1, 3 and 10 $\mu$eV from bottom to top for $p$-InAs 
and is fixed at 0.2 $\mu$eV for $p$-InSb.  Other material parameters
including $\Delta_{exc}$ for GaAs and GaSb are as specified in Table
I.
}
\end{figure*}
\noindent
\pagebreak
\begin{figure*}
\centerline{\epsfxsize=14cm \epsfbox{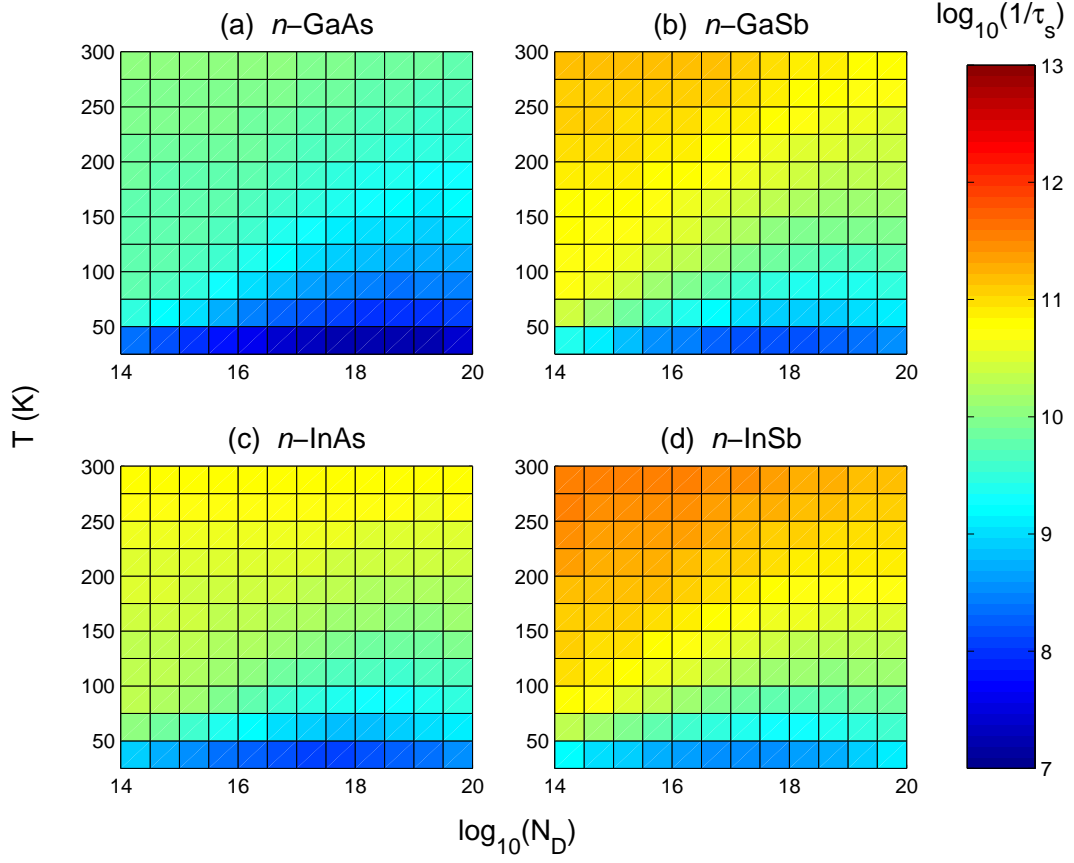}}
\vspace{4mm}
\caption{Total spin relaxation time for $n$-type materials.  $N_D$ is
in cm$^{-3}$ and $\tau_s$ is in second.  $N_A$ is fixed to $5\times10^{13}$ 
cm$^{-3}$.
}
\end{figure*}
\noindent
\pagebreak
\begin{figure*}
\centerline{\epsfxsize=14cm \epsfbox{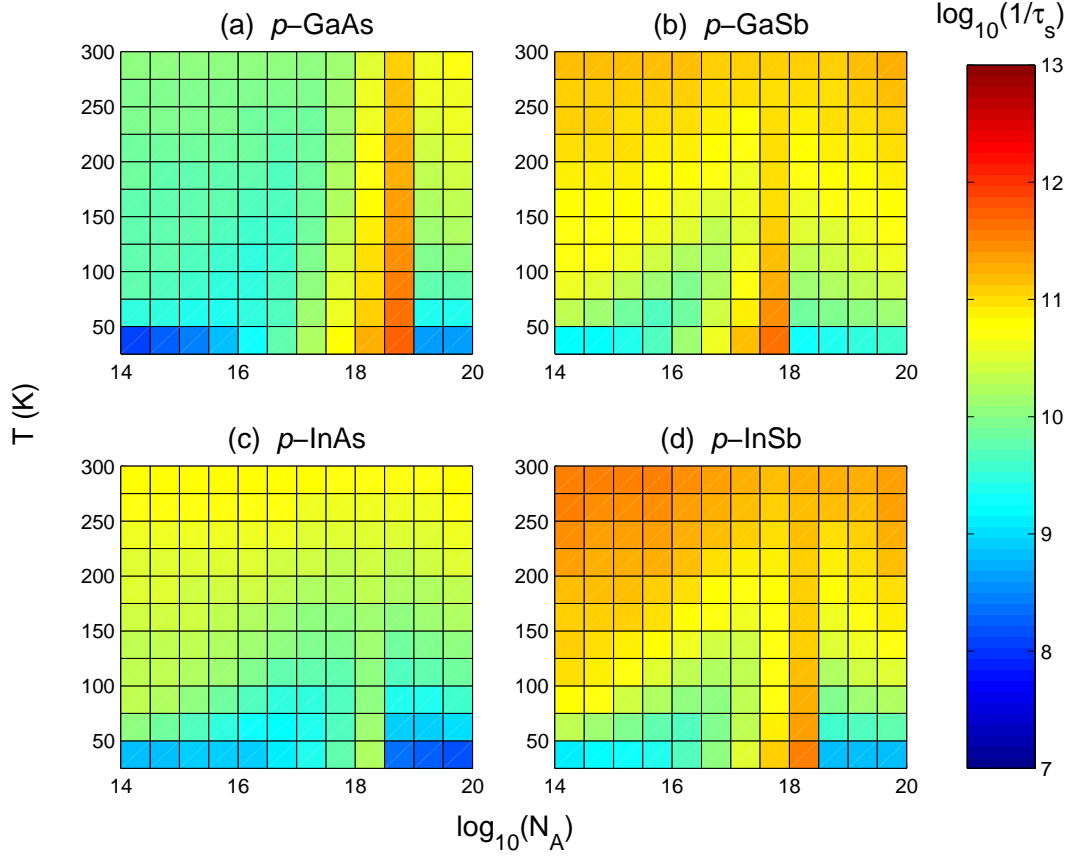}}
\vspace{4mm}
\caption{Total spin relaxation time for $p$-type materials.  $N_A$ is
in cm$^{-3}$ and $\tau_s$ is in second.  $N_D$ is
fixed to $5\times10^{13}$ cm$^{-3}$ and $\Delta_{exc}$ to 
1 $\mu$eV for $p$-InAs and $p$-InSb.
}
\end{figure*}
\noindent

\end{document}